\documentclass[12pt,preprint]{aastex}

\slugcomment{Received 2007 September 14; accepted 2007 December 5}

\shorttitle{Rich Molecular Gas in Two Narrow-Line Seyfert 1 Galaxies}
\shortauthors{KAWAGUCHI et al.}

\def\brfrac#1#2{\left(\dfrac{#1}{#2}\right)}
\def\dfrac#1#2{{\displaystyle\frac{\mathstrut #1}{#2}}}
\def\ea{{\rm et~al.\ }}

\def\ltsima{$\; \buildrel < \over \sim \;$}
\def\lesssim{\lower.5ex\hbox{\ltsima}}
\def\Mdot{{\dot{M}}}

\def\Mbh{{M_{\rm BH}}}
\def\Mbulge{{M_{\rm bulge}}}
\def\MH2{{M(H_2)}}
\def\Msun{{M_\odot}}

\def\Lsun{{L_\odot}}
\def\Ledd{{L_{\rm Edd}}}

\def\Hb{{{\rm H} \beta}}
\def\Fe2{{{\rm Fe} \sc II}}

\begin{document}
\title{
FIRST \ DETECTION \ OF \ ${}^{12}$CO ($1 \rightarrow 0$) \ EMISSION \ 
FROM \ TWO \ NARROW-LINE \ SEYFERT \ 1 \ GALAXIES}

\author{Toshihiro Kawaguchi\altaffilmark{1}, 
Kouichiro Nakanishi\altaffilmark{2},
Kohtaro Kohno\altaffilmark{3}, 
Kouji Ohta\altaffilmark{4}
{\rm and} Kentaro Aoki\altaffilmark{5}}

\email{kawaguti@phys.aoyama.ac.jp}

\altaffiltext{1}{Department of Physics and Mathematics, 
  Aoyama Gakuin University, 
  Sagamihara, Kanagawa 229-8558, 
  Japan}
\altaffiltext{2}{Nobeyama Radio Observatory, Minamimaki, Minamisaku, 
  Nagano 384-1305, Japan}
\altaffiltext{3}{Institute of Astronomy, The University of Tokyo,
  Mitaka, Tokyo 181-0015, Japan}
\altaffiltext{4}{Department of Astronomy, Kyoto University, 
  Kyoto 606-8502, Japan}
\altaffiltext{5}{Subaru Telescope, National Astronomical Observatory 
  of Japan, 650 North A'ohoku Place, Hilo, HI 96720, USA}

% For emulateapj
%\centerline{e-mail: kawaguti@phys.aoyama.ac.jp}
% For submit (aastex)

\begin{abstract}
In order to investigate how the growth of galactic bulges 
progresses with the growth of central black holes (BHs), 
we observed molecular gas 
(fuel for the coming star formation) 
in possibly young active galaxies, 
narrow-line Seyfert 1 galaxies (NLS1s).
We present the results of 
radio observations of ${}^{12}$CO\,($1 \rightarrow 0$) 
using the Nobeyama Millimeter Array (with 2--4 kpc spatial resolution) 
for two FIR-bright NLS1s, 
yielding the first detection of their CO emission. 
Corresponding molecular--gas masses $\MH2$ of $(1-3) \times 10^9 \Msun$ 
are the 2nd and 4th largest ones among NLS1s.
By estimating dynamical masses and bulge masses $\Mbulge$ 
 for these two NLS1s using 
 CO channel map and CO line widths, 
 we found $\MH2$ amount to 0.13--0.35 %0.45 
of these masses.
Taking account the star formation efficiency ($\sim 0.1$), 
the increase in $M_{\rm bulge}$ in those NLS1s in the near future 
($\lesssim 10^{7.5}$\,yr) is expected not to be a huge fraction 
(1--5\,\% %8\,\% 
of the preexisting stars).
Bulge growth may have finished before BH growth, or 
bulge--BH coevolution may 
proceed with many, occasional discrete 
events, where one coevolution event produces only a small amount
of mass growth of BHs and of bulges.
We also discuss the ratios of 
star-formation rate--to--gas accretion rate onto BHs,
finding that two NLS1s have very small ratios 
($\approx$\,1) 
compared with 
 the $\Mbulge / \Mbh$ ratios found in active and inactive
 galaxies ($\approx$\,700).
This huge difference 
suggests either 
the non-overlapped coevolution, long star formation duration
or temporarily high accretion rate during NLS1 phase.
\end{abstract}


\section{INTRODUCTION}

%=== Coevolution ===\\
Mass of galactic spheroid component 
(galactic bulge or elliptical galaxy itself) is tightly correlated 
with mass of central black hole (BH) both in 
normal and active galaxies 
(Gebhardt et al.\ 2000; Ferrarese \& Merritt 2000; Nelson 2000).
Along with the lack of its cosmic evolution (Shields et al.\ 2003;
Kiuchi \ea 2006; 
see however, Akiyama 2005, Woo et al.\ 2006, and Borys \ea 2005), 
the tightness of the correlation 
indicates that the growth of spheroid mass 
and that of BH mass ($M_{\rm BH}$)
occur almost contemporarily. 

However, 
no direct evidence for %(quasi-) 
simultaneous coevolution has been found,
and thus its physical origin remains a great mystery. 
Jet and/or outflow 
from an Active Galactic Nucleus (AGN) 
may govern the formation of spheroid (Silk \&
Rees 1998); i.e., BHs made before the bulges' growth.
On the contrary, radiation drag by massive stars during the
bulge formation may have controlled the BH growth
(Umemura 2001), resulting in bulges formed 
prior to BHs' growth.

Most numerical studies presume that cool, interstellar gas 
drives coevolution (e.g., Kauffmann \& Haehnelt 2000).
For instance, semi--analytical approaches 
assume a major merging of galaxies initiates the growth of 
BHs and bulges, and 
usually adopt 
an AGN duration ($e$--folding accretion time scale) around 
$10^{7.5}$\,yr and 
a star formation duration of $\sim 10^8$\,yr 
(Kauffmann, White \& Guiderdoni 1993; Kauffmann \& Haehnelt 2000). 
Theoretically and observationally, it is unknown %highly uncertain 
which growth phase precede another. 
Now, let us simply assume that the 
BH and bulge growth phases begin conjunctionally 
and proceed with similar durations. 
Then, we will see enough 
molecular gas -- fuel for the coming star formation -- 
in the bulges 
of AGNs in the early stage of BH growth 
(e.g., in the first quarter of the AGN/star formation durations).
Old AGNs (e.g., in the last quarter of the durations) 
that have spent long time since the
ignition of coevolution would have already consumed the gas 
through the past star formation, 
with only a tiny amount of leftover for further bulge growth.

%=== About NLS1 ===\\
Here, we notice a subsample of AGNs, 
narrow-line Seyfert 1 galaxy (NLS1),
as possibly young AGNs.
NLS1s are characterized (see Pogge 2000) by 
(i) 
narrower Balmer lines of hydrogen 
(FWHM of $\Hb$ $\leq$ 2000 km s$^{-1}$), relative to
usual broad-line Seyfert 1 galaxies (BLS1s) and QSOs having
FWHM of $\Hb$ $\gtrsim$ 5000 km s$^{-1}$ (Osterbrock \& Pogge 1985).
(ii) They often emit strong optical 
\Fe2 multiplets (e.g., Halpern \& Oke 1987), and show
(iii) 
steep and luminous soft X-ray
excess (Pounds et al. 1995; Otani et al. 1996; Boller et al. 1996; 
Wang \ea 1996; Laor et al. 1997; Leighly 1999b).
(iv) Rapid X-ray variability is another characteristic 
of NLS1s (Otani et al. 1996; Boller et al. 1997; 
Leighly 1999a; Hayashida 2000).
Characteristics in (ii)--(iv) indicate that NLS1s are non-obscured
objects, alike type 2 (i.e., obscured) AGNs.
Currently, the most promising picture of NLS1s 
is that they contain
relatively less massive black holes (with $\Mbh \sim 10^{6-8}M_\odot$) and
higher $L/\Ledd$ ratio ($\sim 1$; e.g., Brandt \& Boller 1998;
 Hayashida 2000; Mineshige \ea 2000),
where $L$ and $\Ledd$ are the AGN luminosity
and Eddington luminosity, respectively.
NLS1s occupy 10--30\% among type 1 AGNs in the local universe 
(see Kawaguchi \ea 2004b).

High $L/\Ledd$ ratios of NLS1s imply that 
their BHs are now growing rapidly via super-Eddington 
accretion (Rees 1992; Mathur 2000; Kawaguchi et al. 2004b).
Provided that a gas accretion rate onto the central BH, 
$\Mdot$, during an AGN phase does not drastically change with time 
(see Collin \& Kawaguchi 2004), 
large $\Mdot / \Mbh$ ratios of NLS1s also mean that 
the elapsed time since the ignition of an AGN activity
is shorter than that of BLS1s.
(If the gas accretion lasts long enough, then $\Mbh$ would not
be so small.) 
Bulge luminosity 
of NLS1s are smaller than those of BLS1s 
in general (Botte et al. 2004).

NLS1s are likely young active galactic nuclei 
[during the first 10--30\% of the whole AGN lifetime 
($\sim 10^8$\,yr; see \S 6)],
 in the process of coevolution of 
supermassive BHs and bulges. 
This is our working hypothesis in this study. We consider that
%Thus 
they are the best and unique targets to explore how the BH-bulge
coevolution proceeds.
If the growth phases of BHs and bulges 
are really overlapped, 
bulges of NLS1s may also be rapidly growing with 
a significant amount of 
molecular gas accumulated in their bulges.
Do they show 
large ratios of molecular-gas to bulge masses, 
compared with BLS1s 
(which are expected to be older than NLS1s, 
based on their lower $\Mdot / \Mbh$ ratios)?
To examine these conjecture, 
we need copious data for numerous NLS1s, 
 collected under the condition that
 the spatial resolution is less than the bulge size
 (typically in the order of 1\,kpc). 

However, 
NLS1s are rare 
%only 10--30\,\% 
among Seyfert 1 galaxies, and 
thus we do not have many NLS1s, whose CO emission are detected,
 in the nearby universe.
Therefore, 
observations to increase the number of CO-detected NLS1s
are fundamental for listing strong CO emitters, with an eye to
follow--up sub--kpc observations on them.
The $\Mbh$--bulge mass relation of NLS1s and its comparison 
with the relation of inactive galaxies are subjects of debate 
(Komossa \& Xu 2007, and references therein).
Our observations would also help to clarify if 
NLS1s are really young AGNs. 

In this paper, we present 
the observational results of 
${}^{12}$CO\,($1 \rightarrow 0$) line from 
2 NLS1s with 5--6'' (2--4 kpc at their distance) spatial resolution
using the Nobeyama Millimeter Array (NMA) at 
the Nobeyama Radio Observatory.
%=== Structure of paper and Parameter ===\\
The next section outlines the targets selection, and 
 \S 3 describes the log and data reduction of our radio observations.
We then show the results in \S 4, 
 followed by discussions on bulge--mass to stellar--mass ratios 
  in \S 5.
The final section is devoted to discussions and summary.
Cosmological parameters adopted in this paper are 
$H_0 = 70$\,km\,s$^{-1}$\,Mpc$^{-1}$,
$\Omega_M = 0.3$ and $\Omega_\Lambda = 0.7$.

\section{Selection of Targets}

So far, there are only a few NLS1s 
whose CO lines are detected in the nearby universe.
Intending to increase the number of such NLS1s, 
we thus carried out the following 
observations 
(with beam size $\sim \,$6'') onto two NLS1s 
with no CO observations reported.

In order to select targets, 
we investigated flux densities
at 100$\mu$m (measured with IRAS satellite), which roughly trace the 
CO line flux (Solomon et al.\ 1997), 
for all NLS1s at Decl.\ $> -25^{\circ}$ and
redshift $z \leq 0.067$ (the sample summarized by Ohta et al.\ 2007).

For NGC~4051, 
of the largest 100$\mu$m flux (24\,Jy), its CO emission line
is clearly detected (Young \ea 1995; Maiolino \ea 1997; 
 Kohno \ea 1999a,\,b).
Whereas only an upper-limit is obtained for the 
second 100$\mu$m brightest (4.7\,Jy) object (Mrk~766 = NGC~4253),
with $M(H_2) < 10^{7.7} \Msun$ (Taniguchi \ea 1990). 
It then turned out that the third and forth brightest 
NLS1s ($\sim$\,4\,Jy) 
have no radio observations made on their molecular gas.
We select these two SBb galaxies, 
IRAS~04312+4008 ($z=$0.020) and
IRAS~05262+4432 ($z=$0.032), to observe.

We regard these two NLS1s as the best targets to elucidate how bulge
growth is accompanied with BH growth:
By assuming a viliarized 
Broad Line Region (Wandel et al.\ 1999; Kaspi et al.\ 2000),
$M_{\rm BH}$ is estimated with H$\beta$ width:
\begin{equation}
 \Mbh = 4.82 \,\times \,10^6 \Msun 
  \left[\dfrac{{\rm FWHM (H}\beta)}{\rm 1000 \, km \,s^{-1}}\right]^{2}
  \left[\dfrac{\nu L_\nu (5100 \AA)}{10^{44} \,
    {\rm erg \, s^{-1}}}\right]^{0.7}.
 \label{eq:bhmass}
\end{equation}
Although some improved equations in place of the equation above 
are suggested by, e.g., Onken \ea (2004), Kaspi \ea (2005) and
Collin \ea (2006), 
we adopt eq.~\ref{eq:bhmass} in this paper for simplicity.
The FWHM(H$\beta$), optical luminosity and estimated BH masses
for the two objects are
%(860\,km\,s$^{-1}$, 10$^{44.62}$\,erg\,s$^{-1}$, 10$^{6.98} \Msun$) 
(860\,km\,s$^{-1}$, 10$^{44.6}$\,erg\,s$^{-1}$, 10$^{7.0} \Msun$) 
and 
%(740\,km\,s$^{-1}$, 10$^{45.22}$\,erg\,s$^{-1}$, 10$^{7.28} \Msun$) 
(740\,km\,s$^{-1}$, 10$^{45.2}$\,erg\,s$^{-1}$, 10$^{7.3} \Msun$) 
for IRAS\,04312 and IRAS\,05262, respectively (Ohta et al.\ 2007). 
Based on the $\Mbh$--estimations in this way 
 and
our spectral models for $\Mdot$--evaluation (Kawaguchi 2003), 
they have extremely small $M_{\rm BH} / \dot{M}$ ratios 
($< \,$2~Myr) even among NLS1s [all CO data for NLS1s
taken from our observations and from literatures 
will be given in the forthcoming paper (Kawaguchi \ea in prep.)].
Thus, they probably have spent little time since each BH began to
grow.

%Some supportive evidences for on-going star formation 
%in the two objects are: 
%(i) Our optical 
%imaging data of their host galaxies (Ohta et al.\ 2007) show
%axis-asymmetric morphology (bars), 
%implying recent star formation activities.
%(ii) IRAS~05262 
%shows an extended narrow line emission with low 
%[N\,II]/H$\alpha$ ratio, i.e.\ from H\,II regions 
%(V\'{e}ron-Cetty \ea 2001).

\section{Observation and Data Reduction}

In April--March 2006,
aperture synthesis observations toward 2 NLS1s were 
carried out with the NMA D-configuration,
with the longest baseline of 82m.
The beam size ($\sim 5'' \times 6''$) corresponds to 
 2--4\,kpc at their distances (see \S\,5 for 
 their bulge effective radii).
The NMA consists of six 10m antennas,
although one antenna was sometimes out of order 
during our observations. 

In all observations, 
\ensuremath{{\hfill12\atop\hfill} \mathrm{CO}} ($1 \rightarrow 0$) 
($\lambda_{\rm rest} = 2.6008$ mm or $\nu_{\rm rest} = 115.271$ GHz)
lines were 
searched for, by setting the central frequency 
of each observation to the redshifted frequency.
The backend we used was the Ultra-Wide-Band Correlator 
(UWBC; Okumura et al. 2000), 
which was set to cover 512MHz ($\sim$1300 km\,s$^{-1}$) 
with 256 channels at 2 MHz resolution 
(with $\sim$\,10\,km/s effective resolution after 
the Hanning window function applied). 
The quasar 
3C111 
was used for phase and amplitude reference calibrations, 
while 3C454.3 was used for bandpass calibration. 
Total on-source integration time on each object
and beam size are summarized in Table~1. %\ref{tab:log}.
To calibrate absolute flux-scale, 
Uranus was used. 
The uncertainty in the absolute flux scale is typically
 $ \sim$\,$10\%$. 
Standard data reduction was performed 
using the UVPROC-II software package developed at NRO
(Tsutsumi et al. 1997).
Some parts of data taken under a bad radio seeing condition were
removed from our analysis.
After clipping a small fraction of data with unusually high-amplitude,
the data were then imaged with natural UV weighting, and
CLEANed, with the NRAO AIPS package.
The absolute positional accuracy is less than $\sim$\,5\,\% 
of the beam size.

\section{Results}

The CO emission from the 2 IRAS galaxies was detected, for the first time,
by our observations.

\subsection{Distribution of Gas}

\begin{figure*}[tb]
% For emulateapj
%\vspace*{0.5cm}
%\figurenum{1}
%\centerline{\includegraphics[angle=0,width=8.5cm]{IR0431-CL2.ICLN.superimposed.eps}}
\plotone{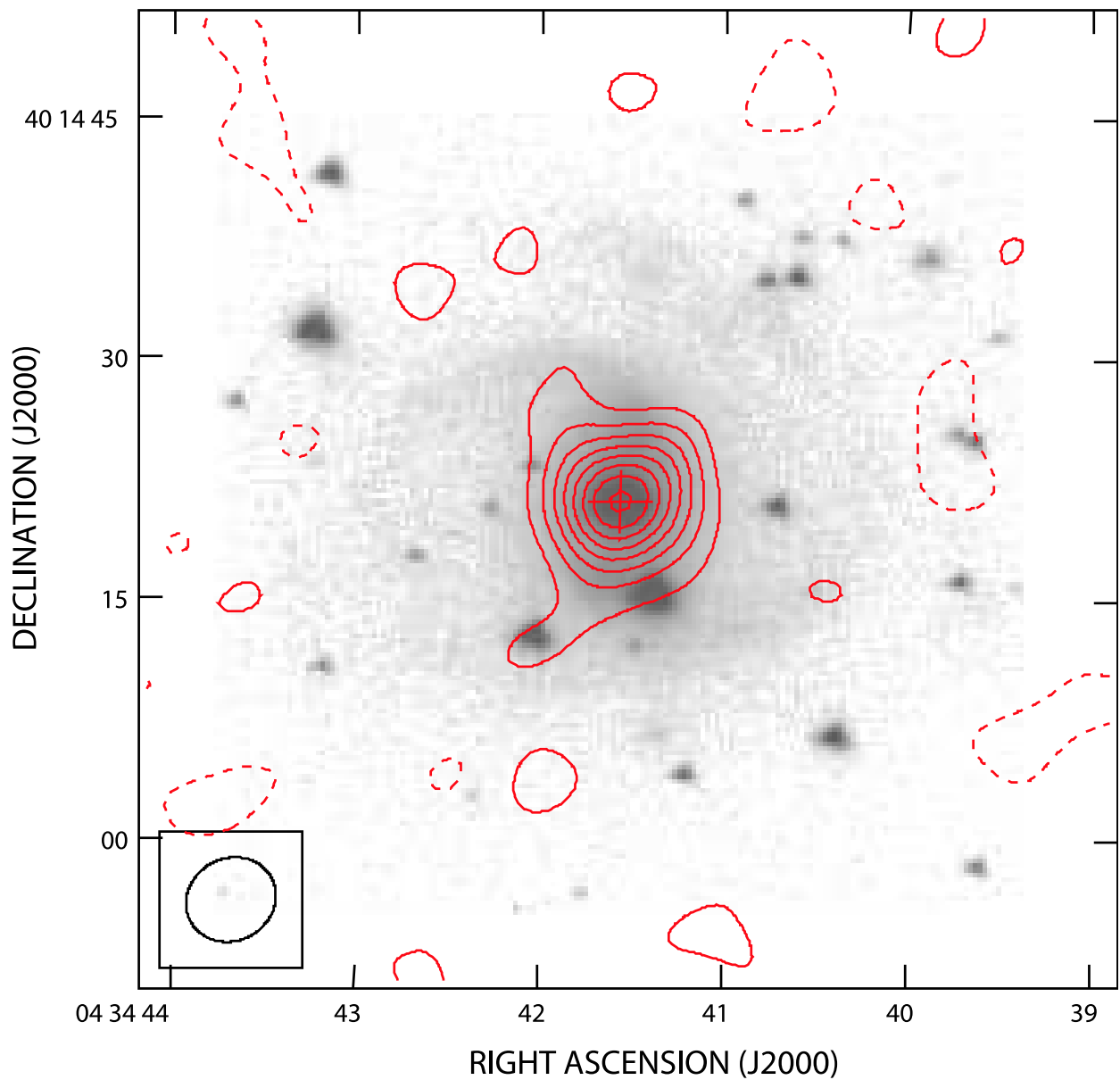}
%\plotone{f1.pdf}
\caption{
CLEANed contour map of CO\,($1 \rightarrow 0$) emission line
from IRAS~04312 binned in frequency over 320\,km\,s$^{-1}$, 
on top of the optical image (Ohta et al.\ 2007). 
Contour levels show $-2$, 2, 4, $\ldots 14, \, 16\ \sigma$, with 
$1 \sigma = 2.0$\,Jy\,beam$^{-1}$\,km\,s$^{-1}$.
The 4''$\times$4'' cross represents the peak position of the CO distribution. 
At the left-bottom corner, the beam size (5.8'' $\times$ 5.2'')
is shown.
\label{fig:ir043binmap}}
\end{figure*}

\begin{figure*}[tb]
% For emulateapj
%\vspace*{0.5cm}
%\figurenum{2}
%\centerline{\includegraphics[angle=0,width=8.5cm]{IR0526-CL2.ICLN.superimposed.eps}}
\plotone{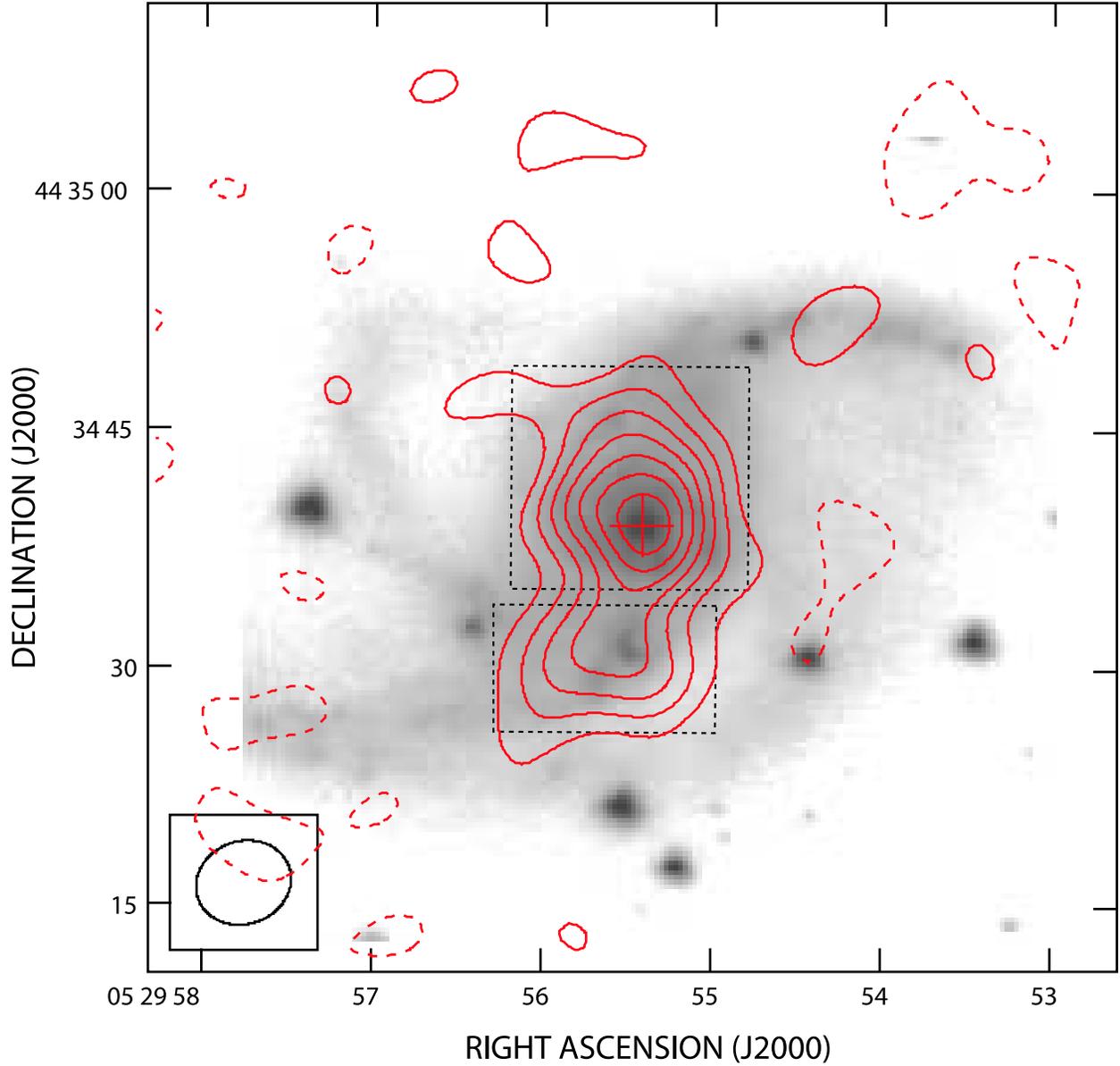}
%\plotone{f2.pdf}
\caption{
The same as figure \ref{fig:ir043binmap}, but for IRAS~05262.
Larger square including the galactic center 
and another square, 
south from the center, 
 indicate the areas over which the
spectra (figure \ref{fig:ir052spec}) are integrated.
The contour map shows the CO emission integrated over
190\,km\,s$^{-1}$, with 
$1 \sigma = 1.2$\,Jy\,beam$^{-1}$\,km\,s$^{-1}$.
The beam size is 6.1'' $\times$ 5.2''.
\label{fig:ir052binmap}}
\end{figure*}

\begin{figure*}[tb]
% For emulateapj
%\vspace*{0.5cm}
%\figurenum{5}
%\centerline{\includegraphics[angle=0,height=6.5cm]{IR043-CL0-R.MCUBE.2.eps}}
\plotone{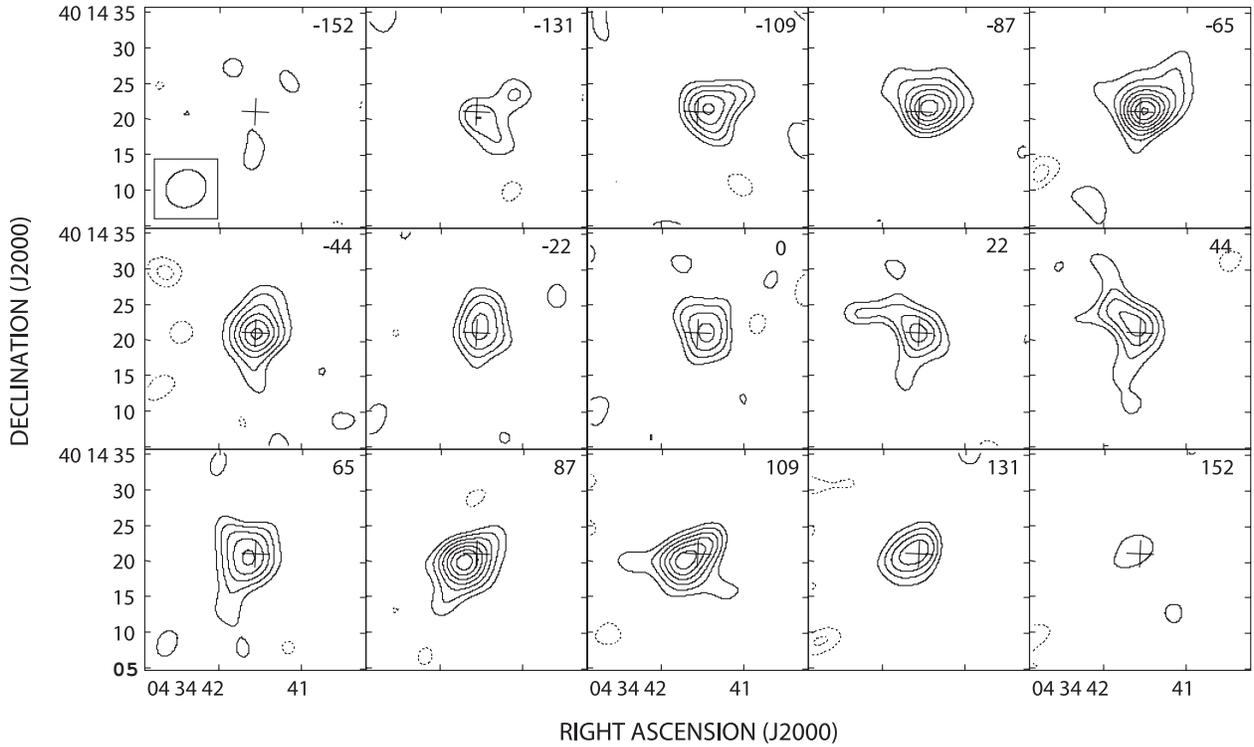}
%\plotone{f5.pdf}
%
\caption{
CLEANed channel map of CO\,($1 \rightarrow 0$) 
emission line around IRAS~04312 over $30'' \times 30''$. 
Each map is binned in frequency over 42\,km\,s$^{-1}$, 
 showing (-3, -2, 2, 3, 4, 5 $\ldots) \times \sigma$
 with $1 \sigma = 19$\,mJy per beam, 
and drawn every 21\,km\,s$^{-1}$ step. 
Crosses represent the peak position of the full CO emission %$\pm 2''$ 
(Fig.\ \ref{fig:ir043binmap}).
Number at upper-right corner of each panel represents the
velocity offset (in km\,s$^{-1}$) from the recession velocity of the galaxy 
(6145\,km\,s$^{-1}$).
\label{fig:ir043chanmap}}
\end{figure*}

\begin{figure*}[tb]
% For emulateapj
%\vspace*{0.5cm}
%\figurenum{6}
%\centerline{\includegraphics[angle=0,height=6.5cm]{IR052-CL0-R.MCUBE.1.eps}}
\plotone{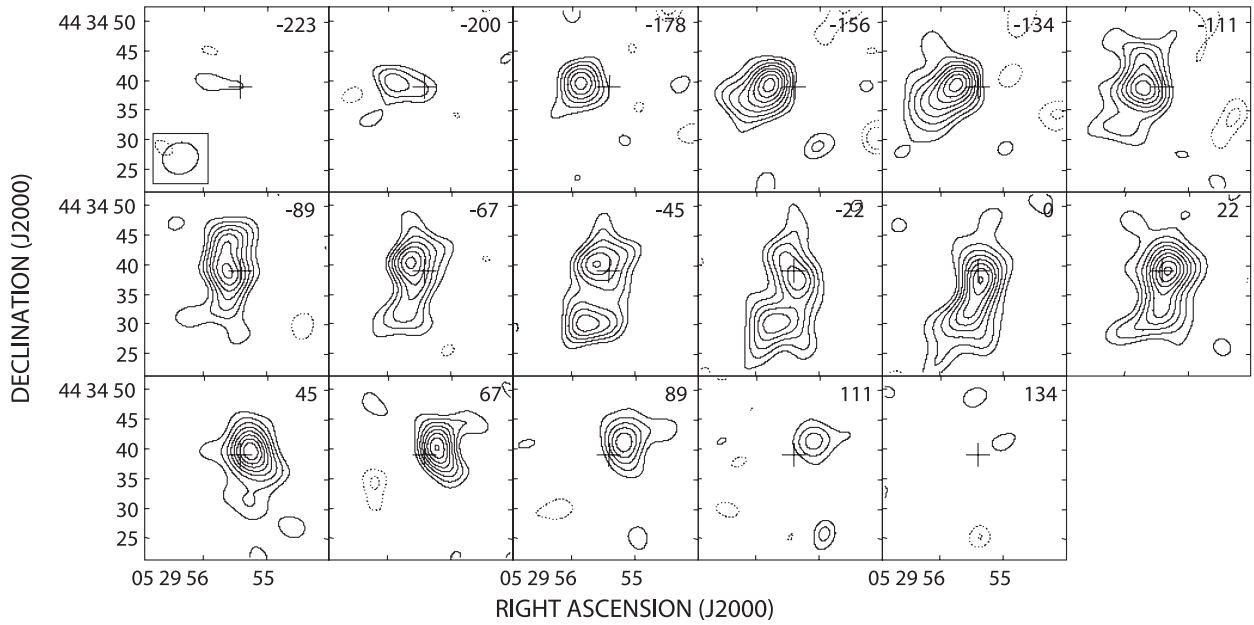}
%\plotone{f6.pdf}
%
\caption{
The same as figure \ref{fig:ir043chanmap}, but for IRAS~05262 
(with a recession velocity of 9646\,km\,s$^{-1}$).
Here, $1 \sigma = 14$\,mJy per beam.
\label{fig:ir052chanmap}}
\end{figure*}

To show the distribution of the molecular gas of the
two targets, we here show the CLEANed maps binned 
over broad velocity ranges (320 or 190\,km\,s$^{-1}$)
in figures \ref{fig:ir043binmap}--\ref{fig:ir052binmap}.
Throughout, north is at the top and east to the left in maps. 
The ranges integrated are shown in 
figures \ref{fig:ir043spec}--\ref{fig:ir052spec} with labels ``map''.
In both objects, the peak flux densities amount up to $\sim 16 \sigma$ 
($\sim 100$\,mJy per beam).
By looking at the 2MASS images, we find that 
the peak of CO emission from each galaxy coincides 
with the peak position of the NIR emission of each galaxy.
The peak position of the CO emission is marked with the 
crosses, to refer the central position in each 
channel map later (figures \ref{fig:ir043chanmap}--\ref{fig:ir052chanmap}).
Optical images taken at University--of--Hawaii 88 inch telescope
at $I$-band, taken under $\sim 0.6''$ (FWHM) image quality 
(Ohta \ea 2007), 
are also shown to present the morphology of the host galaxies.

The gas distribution of IRAS~04312 is quite centrally concentrated
(fig. \ref{fig:ir043binmap}).
The FWHM of this signal is about 7.2'' $\times$ 6.8''.
Considering the beam size (5.8'' $\times$ 5.2''), the 
spatial extent (in diameter) of the gas is estimated to be 
$\sim$ 4.3'' $\times$ 4.4'' ($\sim$\,1.7--1.8\,kpc),
well smaller than the beam.
On the other hand, 
the molecular gas of IRAS~05262 (fig. \ref{fig:ir052binmap}) seems 
extended larger than the beam size (6.1'' $\times$ 5.2'').
The FWHM of its central CO emission, $\sim$\,8.3''$\times$10.7'', 
suggests the actual size to be 
$\sim$\,5.6''$\times$9.4'' (3.5$\times$5.9\,kpc),
elongated in the North--South direction.
In \S\,5, 
these sizes are compared with their bulge effective radii, 
estimated by a couple of methods, 
to discuss how much fraction of the detected CO gas 
are located within the bulge.

Incidentally, an isolated CO emission is seen 
$\sim$\,$7''$ ($\sim$\,4.4~kpc in projected distance) south
from the center of IRAS~05262, which seems to 
arise from a star forming region evident 
in the optical image.

\subsection{CO Spectra and Molecular-Gas Masses}

\begin{figure*}[tb]
% For emulateapj
%\vspace*{0.5cm}
%\figurenum{3}
%\centerline{\includegraphics[angle=0,width=6.5cm]{IR043-CL0-R.MCUBE.2.spec2.eps}}
\plotone{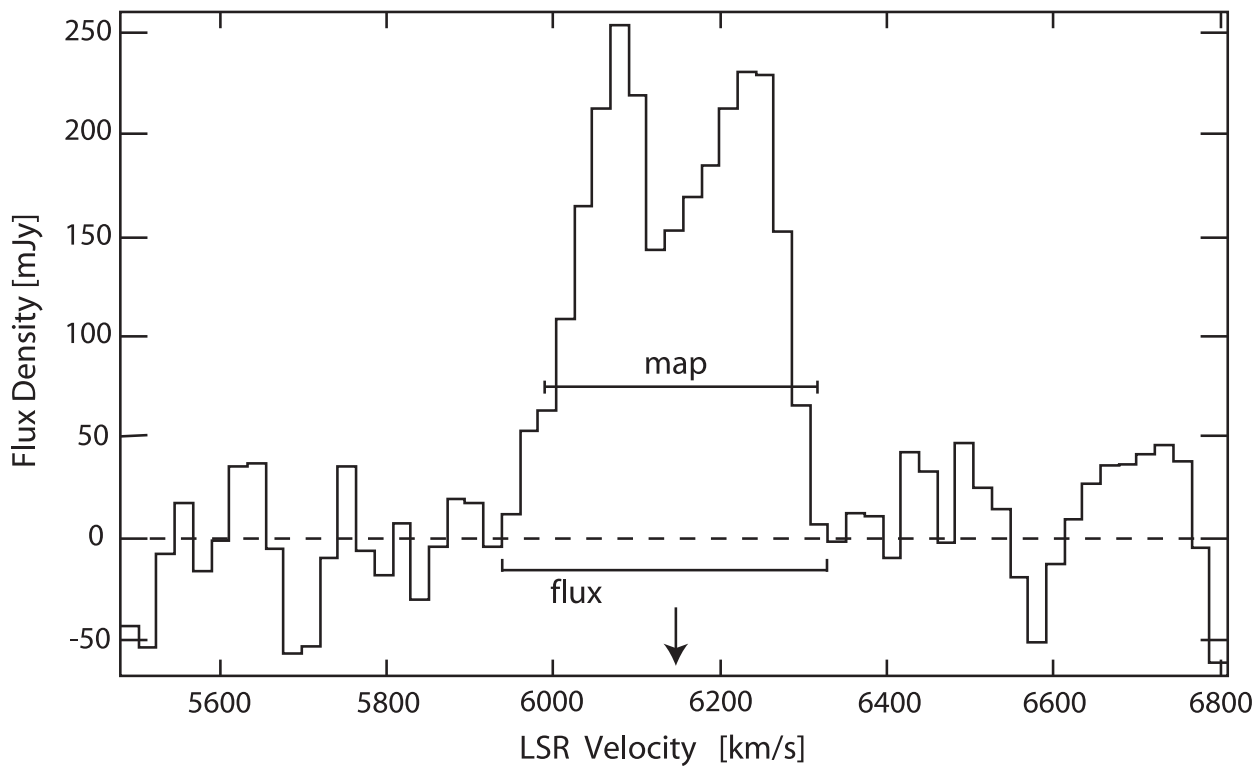}
%\plotone{f3.pdf}
%
\caption{
The CO\,($1 \rightarrow 0$) spectrum
of the central region of IRAS~04312, integrated over 
12''$\times$12'' box centered on
the peak of the CO emission.
(Hereafter, velocity notation is based on optical convention.) 
The bin--width and frequency--step are the same as 
those for the channel maps (fig.\ref{fig:ir043chanmap}).
Downward arrow indicates 
the recession velocity of the galaxy,
measured with H~I gas (Paturel et al. 2003).
The horizontal solid line indicates the range
used in figure \ref{fig:ir043binmap} 
to draw the frequency--integrated map.
Another horizontal line with a label ``flux'' indicates 
the range for measuring the velocity--integrated flux.
\label{fig:ir043spec}}
\end{figure*}

\begin{figure*}[tb]
% For emulateapj
%\vspace*{0.5cm}
%\figurenum{4}
%\centerline{\includegraphics[angle=0,width=6.5cm]{IR052-CL0-R.MCUBE.1.spec.eps}}
\plotone{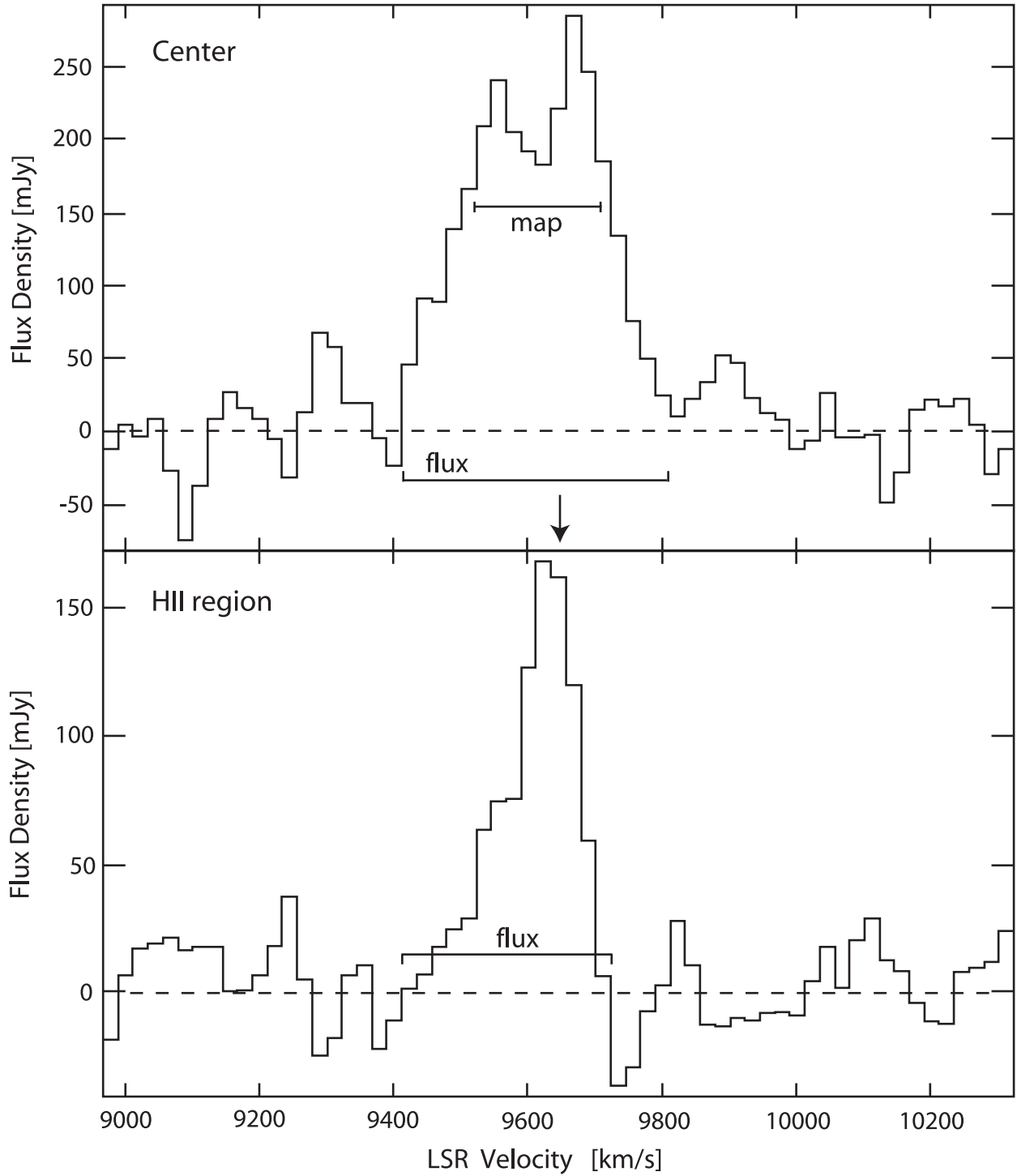}
%\plotone{f4.pdf}
%
\caption{
The same as figure \ref{fig:ir043spec}, but for IRAS~05262.
Top panel shows the spectrum of the central CO gas, while
the bottom presents the spectrum of the 
H~II region $\sim 7''$ south from the center.
\label{fig:ir052spec}}
\end{figure*}

Next, we present spectra of CO\,($1 \rightarrow 0$) line for the 2 NLS1s,
in figures \ref{fig:ir043spec} (IRAS~04312) and 
\ref{fig:ir052spec} (IRAS~05262).
Spectrum of each central region shows a double--horned
shape, indicating a rotating molecular disk in origin.
The horizontal solid lines, labeled as ``map'', show the ranges 
used in figures \ref{fig:ir043binmap}--\ref{fig:ir052binmap}
to draw the frequency--integrated maps.
Downward arrows 
indicate the recession velocity of the targets.

The CO--to--H$_2$ conversion factor 
is estimated to be 0.8 and 2.9 
$M_\odot$ (K km s$^{-1}$ pc$^2$)$^{-1}$ for
ultra-luminous infrared galaxies (Downes \& Solomon 1998)
and for our Galaxy (Dame \ea 2001), respectively.
When converting CO line flux to H$_2$ mass, $\MH2$, 
the conversion factor = 1 $M_\odot$ 
(K km s$^{-1}$ pc$^2$)$^{-1}$ is assumed throughout this paper.
The same conversion factor is applied for
re-computations of 
$\MH2$ for objects taken from literatures.

The velocity--integrated flux from 
IRAS~04312 measured over the same area used 
to draw figure \ref{fig:ir043spec}, is 
57$\pm 3$\,Jy\,\,km\,s$^{-1}$, 
meaning that $M(H_2) = (1.0 \pm 0.05) \times 10^9 \Msun$.
The velocity integration range is shown in figure \ref{fig:ir043spec}
with a label ``flux''.
While, the velocity--integrated flux from the central region of  
IRAS~05262, measured in the box including the CO peak 
(fig.\ref{fig:ir052binmap}), is 
62$\pm 3$\,Jy\,\,km\,s$^{-1}$, 
corresponding to $M(H_2)$ of $(2.8 \pm 0.1) \times 10^9 \Msun$.
These molecular--gas masses are the 2nd and 4th largest masses
(to our best knowledge) among NLS1s,
with I\,Zw\,1 [$\MH2 = 10^{9.5} \Msun$ and $z=0.061$; Maiolino \ea 1997] 
and PG\,1440+356 [$\MH2 = 10^{9.3} \Msun$ and $z=0.079$; Evans \ea 2001] 
being the 1st and 3rd ones.
Summary of all NLS1s whose CO emission are detected
will be given in the forthcoming paper (Kawaguchi et al.\ in prep.).
If we restrict ourselves at $z \leq 0.05$ (where 
$1''$ angular resolution
 is equivalent to 1\,kpc spatial resolution), 
these masses are the largest ones among NLS1s.

In the bottom panel of figure \ref{fig:ir052spec}, 
the spectrum of the isolated CO emission 
($\sim$\,7'' south from the center) is also shown.
This quite narrow emission line has 
a velocity--integrated flux of 21$\pm 1.4$\,Jy\,Km\,s$^{-1}$,
corresponding to $M(H_2) = (9.3 \pm 0.6) \times 10^8 \Msun$.
Translating this $H_2$ mass into a star formation rate (SFR)
by using equations \ref{eq:FIRMH2} and \ref{eq:SFRFIR} shown later,
we get an SFR of 
3.0 $\Msun$\,yr$^{-1}$.

\subsection{Kinematics}

Figures \ref{fig:ir043chanmap} and \ref{fig:ir052chanmap}
show channel maps of the two IRAS galaxies. 
Each map presents a map binned over 42\,km\,s$^{-1}$, and velocity 
step between adjacent maps is 21\,km\,s$^{-1}$.
The cross drawn in each panel represents the peak position
of the full CO emission 
(Figs.\ \ref{fig:ir043binmap} and \ref{fig:ir052binmap}).  
In IRAS~05262, a molecular--gas clump at $\sim 7''$ south from the center 
is evident.

\begin{figure*}[tb]
% For emulateapj
%\vspace*{0.5cm}
%\figurenum{7}
%\centerline{\includegraphics[angle=0,height=6.5cm]{IR043-CL0PV.AVD.eps}}
\plotone{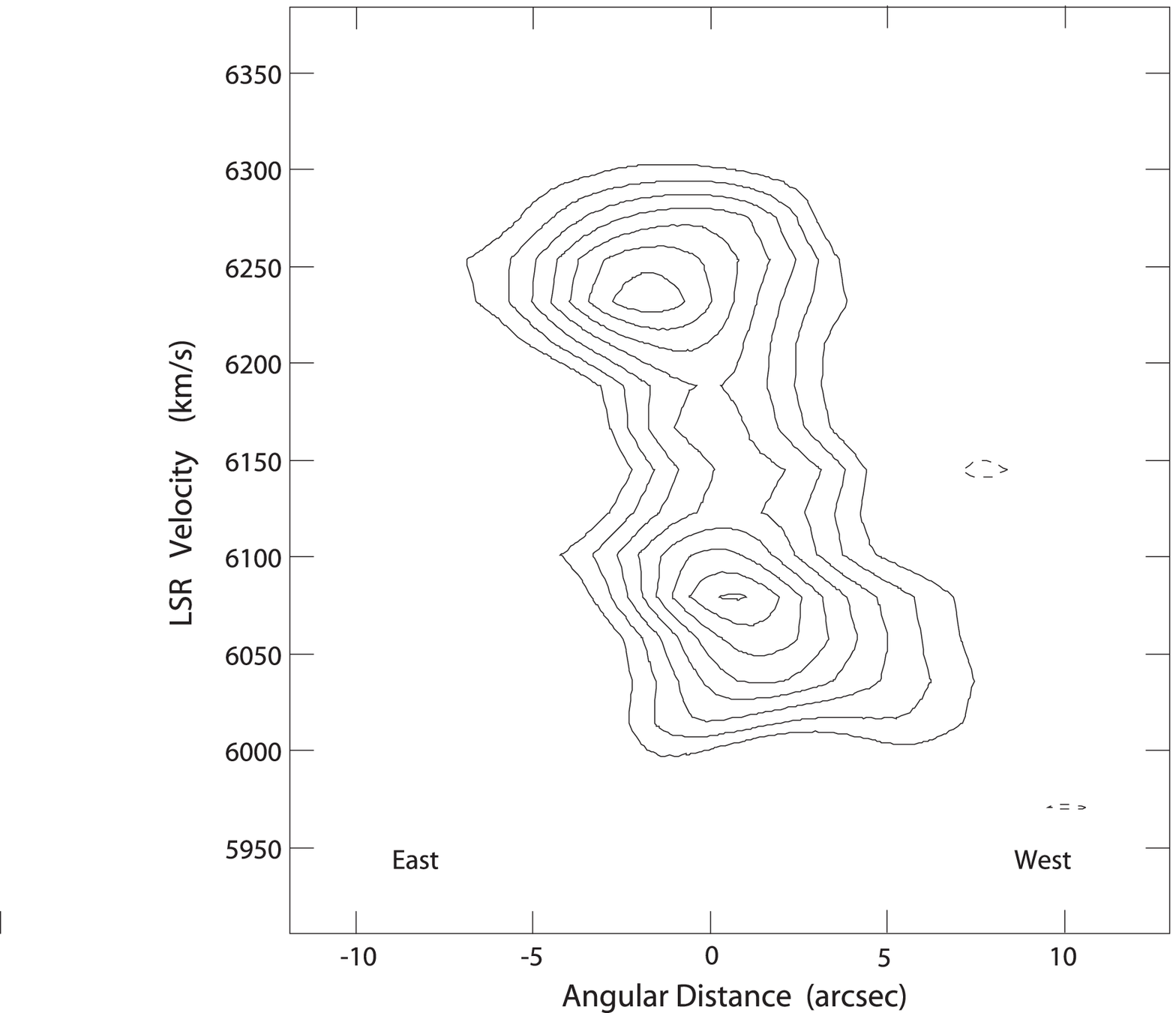}
%\plotone{f7.pdf}
%
\caption{
Position--Velocity diagram of ${}^{12}$CO\,($1 \rightarrow 0$) 
emission from IRAS~04312
along the position angle of 108$^{\circ}$.
The abscissa is the angular distance from the CO emission peak.
The contour levels drawn 
are the same as those of the channel map 
for the same object (fig.\ref{fig:ir043chanmap}).
\label{fig:ir043pv}}
\end{figure*}

\begin{figure*}[tb]
% For emulateapj
%\vspace*{0.5cm}
%\figurenum{8}
%\centerline{\includegraphics[angle=0,height=6.5cm]{IR052-CL0PV.AVD.eps}}
\plotone{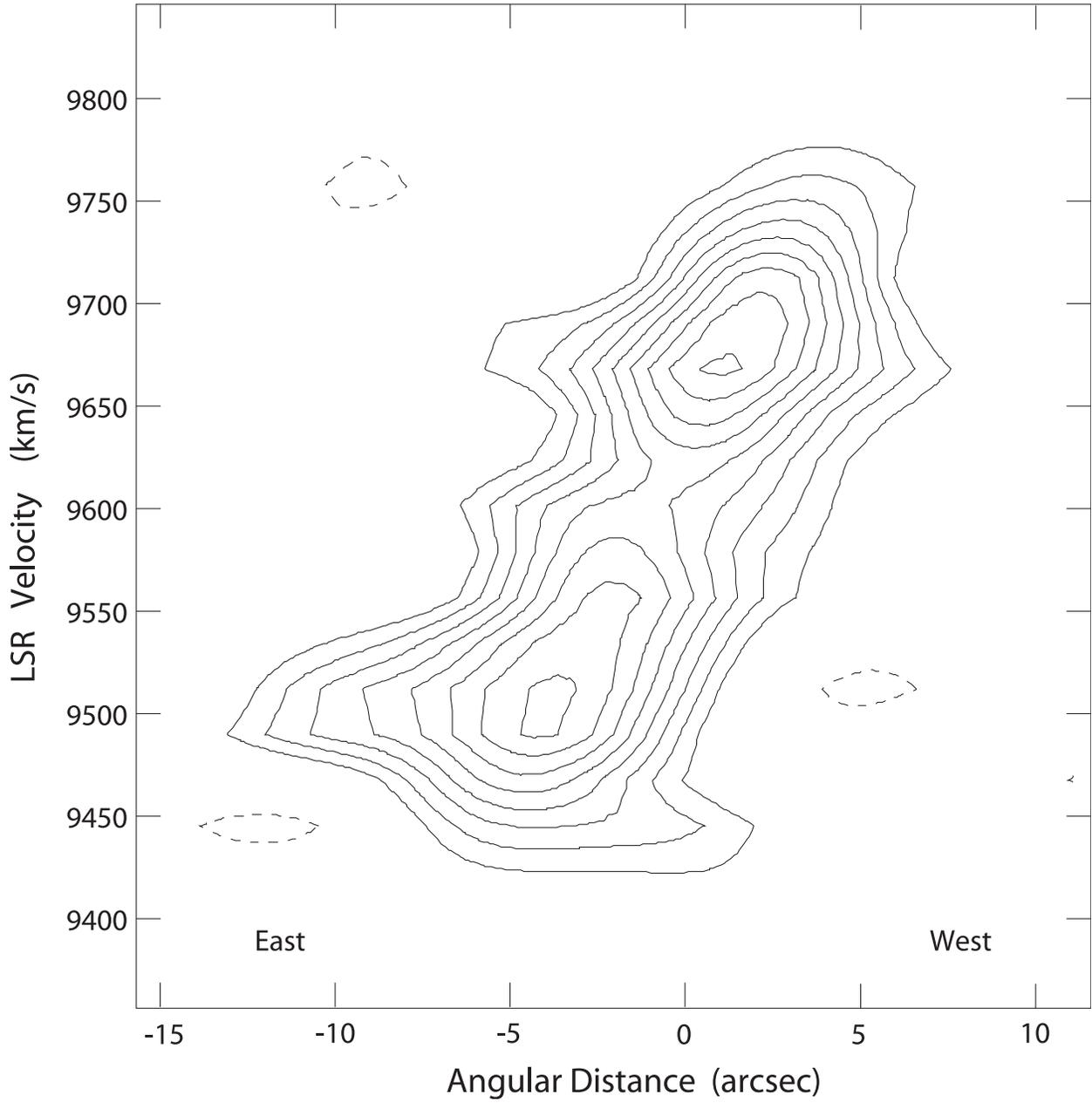}
%\plotone{f8.pdf}
%
\caption{
The same as figure \ref{fig:ir043pv}, but for IRAS~05262 
(with the position angle of 97$^{\circ}$).
\label{fig:ir052pv}}
\end{figure*}

In both objects, rotating dynamics around each center (cross) 
is clearly seen.
Position--velocity diagrams shown 
in figures \ref{fig:ir043pv}--\ref{fig:ir052pv} also indicate
rotation. 
The position angle used for each 
diagram is determined as follows.
Two maps (of high S/N peaks) are chosen from each channel map. 
Then, the peak positions in the two maps 
(with two channels separated by $\sim 210$\,km\,s$^{-1}$)
are connected by a line. 

Now, we derive dynamical mass (i.e., enclosed mass) $M_{\rm dyn}$
from these position--velocity diagrams. 
We estimate an inclination angle $i$ for each galaxy by measuring 
the ellipticity of the {\it I}--band image (Ohta \ea 2007), 
%assume of 60 deg, 
where the inclination angle of 0 deg means face-on view on
a rotating CO gas disk: 
$i=40^{\circ} \pm 1^{\circ}$ (IRAS\,04312) and 
$i=53^{\circ} \pm 5^{\circ}$ (IRAS\,05262). 
Then, the dynamical mass is evaluated as 
\begin{equation}
 M_{\rm dyn} = 5.8 %7.8 
  \times 10^8 \Msun \brfrac{\Delta V}{\rm 100\,km\,s^{-1}}^2 
  \brfrac{R}{\rm kpc} (\sin i)^{-2}, 
%\brfrac{\sin i}{\sqrt{3}/2}^{-2},
\end{equation}
where $\Delta V$ and $R$ are the velocity difference and 
the separation. 
Positional separations $R$ and velocity differences $\Delta V$
between two peaks in the position--velocity diagrams
are 2.7'' (1.1\,kpc in projected distance) and 164\,km\,s$^{-1}$ 
for IRAS~04312, and 
5.1'' (3.2\,kpc) and 165\,km\,s$^{-1}$ for IRAS~05262, respectively.
The enclosed mass $M_{\rm dyn}$ within the central region of IRAS~04312 
is $4.1 \times 10^9 \Msun$. %$2.3 \times 10^9 \Msun$. 
While for IRAS~05262, 
$M_{\rm dyn} = 8.0 \times 10^{9} \Msun$. 
%$M_{\rm dyn} = 6.8 \times 10^{9} \Msun$. 
The peak-to-peak separations 
represent the lower limits of their CO--gas sizes.
As we saw in \S\,4.1, the CO gas in IRAS~04312 is indeed 
more concentrated than that in IRAS~05262.

\section{Sizes of CO Gas and Bulge \label{sec:size}}

We here discuss the sizes of the CO gas in two NLS1s (\S\,4.1 and \S\,4.3),
and compare them with their bulge sizes. 
The spatial extent (in diameter) of the CO gas in IRAS\,04312 
is estimated to be 
$\sim$\,4.3''$\times$4.4'' ($\sim$\,1.7--1.8\,kpc). 
On the other hand, 
IRAS\,05262 seems to have an extended and elongated CO gas at its center 
with the size of 
$\sim$\,5.6''$\times$9.4'' (3.5$\times$5.9\,kpc). 
Peak-to-peak separations 
in position--velocity diagrams 
(Figs \ref{fig:ir043pv} and \ref{fig:ir052pv}), 
2.7'' (=\,1.1\,kpc) and 5.1'' (=\,3.2\,kpc)
 for IRAS~04312 and IRAS~05262, respectively, also indicate
that the molecular gas of IRAS~05262 is more extended than
that of IRAS~04312.

Now, let us estimate the bulge effective radius ($R_{\rm eff}$) 
from [O III] width as follows.
The FWHM of [O III] emission lines of IRAS~04312 
is 380\,km\,s$^{-1}$ (V\'{e}ron-Cetty \ea 2001).
Since [O III] widths and velocity dispersions ($\sigma_*$) are 
nearly the same (Heckman \ea 1989; Nelson \& Whittle 1996), 
$\sigma_*$ for this object is evaluated to be about 160\,km\,s$^{-1}$, 
by dividing the [O III] FWHM width by 2.35.
Elliptical galaxies and bulges of disk galaxies obey
a canonical relation about their dynamics, size, etc (the fundamental plane).
The velocity dispersion can be translated into the
size (Guzman \ea 1993):
\begin{equation}
 R_{\rm eff} = 0.56 \,\,{\rm kpc} 
  \brfrac{\sigma_*}{\rm 100 \,km\,s^{-1}}^{3.15}
\label{eq:Reff}
\end{equation}
Thus, the bulge effective radius of IRAS~04312 is estimated to be 
about 2.6~kpc or 6.4'' at its distance.
However, the inferred bulge size (13'' in diameter) 
is comparable to the length of the bar seen in the 
optical image in the North--South direction.
Therefore, this method likely overestimates $R_{\rm eff}$ for 
this object.
Adopting the same method to IRAS~05262 
by taking its [O III] width (365\,km\,s$^{-1}$) 
 from V\'{e}ron-Cetty \ea (2001), 
its $R_{\rm eff}$ is about
2.3~kpc or 3.6'' at its distance.

We note that there is a large scatter 
about the regression equation (eq.~\ref{eq:Reff}). 
Thus, $R_{\rm eff}$ estimation using eq.~\ref{eq:Reff} alone 
should not be taken too much seriously.
For instance, 
our visual investigations on the radial profile
of the two IRAS galaxies (data presented by Ohta \ea 2007) 
lead us rough estimations of their $R_{\rm eff}$ to be
$\sim 0.7''$ and $3.1''$ for IRAS~04312 and IRAS~05262, 
respectively.
Because of strong emission from the nuclei, 
a large uncertainty of $R_{\rm eff}$ evaluated by 
this method is also inevitable.

For IRAS~04312, two methods above provide so different estimations
(13'' and 1.4'' in diameter).
Thus, it is premature to compare these with the size of the CO gas 
($\sim 4'' \times 4''$), 
and to discuss how much fraction of the total CO flux comes
from its bulge region.
While, two values are consistent each other for IRAS~05262
 (7'' and 6'' in diameter).
Since the spatial extent of the detected molecular gas
 in IRAS~05262 ($\sim$\,5.6''$\times$9.4'') 
 seems a bit larger than the estimated bulge size, 
the molecular gas harbored within the bugle would be
slightly smaller than the detected $\MH2$.
To the first order, the observed $\MH2$ well represents 
the $\MH2$ of the bulge of IRAS\,05262.
It is unclear whether stars newly born from the 
molecular gas belong to the bulge or to the galactic disk. 
In this study, we regard star formation within $R_{\rm eff}$ as 
bulge growth.

\section{Molecular Gas and Bulge Masses}

Here, we compare the molecular--gas masses of these two IRAS galaxies
 with the dynamical masses ($M_{\rm dyn}$) and 
 with the bulge masses ($M_{\rm bulge}$).
These masses are summarized in Table~2. %\ref{tab:mass}.

The dynamical mass $M_{\rm dyn}$ is inferred
from the CO--gas kinematics.
We discuss the position--velocity diagrams 
(figures \ref{fig:ir043pv} and \ref{fig:ir052pv}).
The enclosed mass $M_{\rm dyn}$ within the central region of IRAS~04312 
is $4.1 \times 10^9 \Msun$, resulting in 
a $M(H_2)/M_{\rm dyn}$ ratio of 0.25.
%is $2.3 \times 10^9 \Msun$, resulting in 
%a $M(H_2)/M_{\rm dyn}$ ratio of 0.45.
While for IRAS~05262, 
$M_{\rm dyn} = 8.0 \times 10^{9} \Msun$, and 
the $M(H_2)/M_{\rm dyn}$ ratio is then 0.35.
%$M_{\rm dyn} = 6.8 \times 10^{9} \Msun$, and 
%the $M(H_2)/M_{\rm dyn}$ ratio is then 0.41.
If we estimate the current stellar mass as 
$M_{\rm dyn} - M(H_2)$, 
the molecular gas masses for these two objects account to 
33\,\% and 54\,\% %81\,\% and 69\,\% 
of the current stellar 
masses\footnote{H\,I gas is normally negligible (compared to 
H$_2$ masses) within several kpc (``molecular front'') 
around galactic centers 
(Sofue 1994; Sofue \& Nakai 1994).}.
However, all the molecular gas will not be consumed 
by star formation during this star formation episode 
(in $\sim 10^{7-8}$\,yr), 
since a molecular cloud is to be destroyed by photoevapolation,
mass loss and supernova explosion (Williams \& McKee 1997).
Recycling of the remaining diffuse gas 
%, which is converted into atomic and ionized forms, 
may occur, 
%after efficient radiative cooling, 
but on a timescale much longer than the timescale we 
are considering.
Since we are interested in the BH--bulge evolution during 
one coevolution episode, we dot not go further upon the 
recycling process (on a long timescale) in this study. 
The fraction of a molecular gas mass that is converted 
into stars in molecular clouds (efficiency of star formation) 
ranges from a few percent (Myers et al.\ 1986; 
Wilson \& Matthews 1995) to tens percent (Hillenbrand \& Hartmann 1998; 
Lada \& Lada 2003). 
Adopting a star formation efficiency of 0.1, 
new stars with masses of $3-5 \%$ %$7-8 \%$ 
of the preexisting stars will be formed in the near future 
(e.g., within the timescale of cloud--destruction, 
$\sim 10^{7.5}$\,yr; Williams \& McKee 1997) 
from the molecular gas.

As we discussed in \S 5, the bulge size is not
definitively determined. 
%Due to the uncertainty on the bulge size (\S~5), 
%Since 
Thus, we can not be completely sure that the dynamical mass 
(or virial mass) equals to the bulge mass. 
Therefore, we next try to have another 
estimation of the bulge mass as a complementary measure. 
The bulge mass, $M_{\rm bulge}$, can be estimated from kinematics
of stars/clouds embedded in the gravitational potential of
the bulge, in the form where $M_{\rm bulge} \propto ({\rm line \ width})^4$.
To determine the normalization in this relationship, 
we refer to the $M-\sigma_*$ relationship found in normal galaxies.
If we adopt the two relations, 
$M_{\rm BH} = 8 \times 10^6 \, \Msun 
 \, [{\rm \sigma_*} / ({\rm 100\,km\,s^{-1}})]^4$ 
and 
$M_{\rm bulge} =710 \, \Mbh$ 
(e.g., Tremaine \ea 2002; Gebhardt \ea 2000; H\"{a}ring \& Rix 2004),
we obtain the following relationship:
\begin{equation}
 M_{\rm bulge} = 5.7 \times 10^{9} \Msun 
  \brfrac{\sigma_*}{\rm 100\,km\,s^{-1}}^4.
\label{eq:bulmass}
\end{equation}
The scatter in $\Mbh$--$\sigma_*$ relation is about 0.3\,dex 
(Gebhardt \ea 2000), 
whereas the uncertainty in the $\Mbulge/\Mbh$ ratio is also 
$\sim$\,0.3\,dex (e.g., Marconi \& Hunt 2003).
The uncertainty of the bulge masses evaluated in this way 
would thus be about 0.4\,dex. % (e.g., Merritt \& Ferrarese 2001).

CO widths might be useful to evaluate $\sigma_*$ 
 and then bulge masses 
(Shields \ea 2006; Ho 2007; see \S~\ref{COwidth}).
We here follow their procedure. 
Assuming that the dispersion of CO emission line 
$\sigma_{\rm CO}$ equals to 
$\sigma_*$, as suggested by Shields et al., 
we adopt 
$\sigma_{\rm CO}$ of 109 and 110\,km\,s$^{-1}$ 
for IRAS\,04312 and IRAS\,05262, respectively (see \S\ref{COwidth}).
With eq.~\ref{eq:bulmass}, 
bulge masses are estimated as 
$\Mbulge = 8.0 \times 10^9 \Msun$ and 
$8.4 \times 10^9 \Msun$, 
resulting in $\MH2 / \Mbulge$ ratios of 0.13 and 0.33, respectively.
[If [O III] widths are used as surrogate for $\sigma_*$ 
in eq.~\ref{eq:bulmass}, 
$\MH2 / \Mbulge$ ratios decrease by a factor of 4 (see \S~7.2).] 
%
%Thus, 
Star formation will undergo with %$1.3-3 \%$ 
$1-3 \%$ 
of the masses 
of the current stellar masses for the star formation efficiency of 0.1.
Even if we again estimate the mass of preexisting stars by 
$\Mbulge - M(H_2)$ to account for the contribution of gas mass, 
these numbers do not change dramatically. 

Let us now summarize this section.
Mass of new stars that will be formed 
(within $\sim 10^{7.5}$\,yr from now on) 
from the molecular gas
is not a huge fraction of the current stellar masses (1--5\,\%). 
%(1--8\,\%). 
Here, we conclude that the bulges of these two NLS1s
do not evolve so much in this timescale. 
If newly born stars are to belong to the galactic disks 
rather than the bulges, the conclusion (small growth in the bulge mass) 
is enhanced. 

This result can be interpreted in some ways.
(1) Coevolution is not simultaneous at all:
bulges (and elliptical galaxies) were formed 
in the early universe (Akiyama 2005; Woo \ea 2006), 
while BH growth via accretion still happens today.
%Nevertheless, 
(2) Alternatively, 
coevolution of BHs and bulges can be 
(quasi-) simultaneous with two growth episodes overlapped 
or close in time. 
(2-1) %If this is the case, 
Our result indicates that the 
bulge growth have finished prior to the BH growth (e.g., Umemura 2001). 
(2-2) Or, coevolution may occur 
intermittently through a number of episodic events.
Each coevolving AGN 
could have very short durations of BH and bulge growth, with small
increases in $M_{\rm bulge}$ and in $\Mbh$, 
resulting in a small fraction of molecular gas 
associated with the AGN even in the early evolutionary stage 
(NLS1s).
The total lifetime of an AGN is estimated as $\sim 10^8$~yr 
(Martini 2004; Jakobsen et al. 2003).
While, a lower limit for the duration of an episodic event of an AGN
is about $10^4$ yr, based on the proximity effect (e.g., Bajtlik
et al. 1988) and the sizes of ionization-bounded
narrow-line regions (Bennert et al. 2002).
Therefore, BH--bulge coevolution
could be divided into 
%, at most, $1000$ 
numerous shorter events.

\section{Discussions and Summary}

\subsection{Star Formation Rates v.s. Accretion Rates\label{sec:SFRMdot}}

Here, we discuss the relation between star formation rates
and accretion rates, and the $\Mbulge - \Mbh$ relation.

From FIR luminosity or from molecular--gas mass, 
the current star formation rate can be estimated.
The regression relation between 40--120$\mu$m FIR 
luminosity\footnote{$L_{\rm FIR} = 0.65 \times\nu L_\nu(60 \mu {\rm m})
 + 0.42 \times\nu L_\nu(100 \mu {\rm m})$ (Helou \ea 1985).} 
and $\MH2$ is 
typically (for galaxies similar to the Milky Way) 
expressed as (see Kennicutt 1998; Tinney \ea 1990),
\begin{equation}
 L_{\rm FIR} = 19 \,\Lsun \left[\frac{\MH2}{\Msun}\right].
 \label{eq:FIRMH2}
\end{equation}
The scatter around this linear relation is about 
$\pm 0.3$\,dex (Tinney \ea 1990). 
(Again, the regression is re-normalized here by adopting the 
same conversion factor as we used so far.)
Now, we estimate the FIR luminosity at each central region 
(free from the contribution of AGNs) 
for the two NLS1s, based on their molecular gas masses.
(On the other hand, observed FIR luminosity by IRAS, shown below, 
 includes the FIR emission from entire galaxy and from AGN.) 
Inferred masses (\S\,4.2) indicate that their $L_{\rm FIR}$ 
are $1.9 \times 10^{10}$ $\Lsun$ 
and $5.3 \times 10^{10}$ $\Lsun$, respectively.

By translating these FIR luminosities 
to the star formation rate (SFR) by (Kennicutt 1998)
\begin{equation}
 {\rm SFR} = 1.7 \,\Msun \,{\rm yr}^{-1} \brfrac{L_{\rm FIR}}{10^{10} L_\odot},
 \label{eq:SFRFIR}
\end{equation}
their SFRs (at their central region) 
are expected to be 3.3 and 9.2 $\Msun \,{\rm yr}^{-1}$, respectively.
Most of normalizations for eq.~\ref{eq:SFRFIR} 
in literatures lie within $\pm\,30$\,\% for starbursts.
Disk galaxies, on average, may have a normalization a 
factor of 1.8 higher than eq.~\ref{eq:SFRFIR} (Buat \& Xu 1996).
Taking all the uncertainties of normalizations in eqs.~\ref{eq:FIRMH2} 
and ~\ref{eq:SFRFIR}, an error of SFR estimated from $\MH2$ would be 
$\pm 0.3+0.3$\,dex. 

Incidentally, observed $L_{\rm FIR}$ 
of the 2 IRAS galaxies (measured via IRAS, 
including contributions from entire host galaxy 
and from nuclei) are 
2.7 $\times$ 10$^{10} \, L_\odot$ (IRAS~04312) and 
6.3 $\times$ 10$^{10} \, L_\odot$ (IRAS~05262).
If these $L_{\rm FIR}$ are used instead, 
their SFRs are evaluated to be 5 and 11 $\Msun {\rm yr}^{-1}$, respectively.
Since these values likely overestimate the central $L_{\rm FIR}$ and 
thus central SFRs, 
we hereafter refer the SFRs estimated in the previous paragraph.

Next, we evaluate the accretion rates onto their central BHs $\Mdot$,
by adopting 
two simplified methods. 
Firstly, 
we estimate $\Mdot$ based on bolometric luminosities, 
with an assumption of 
a constant radiation efficiency.
Bolometric luminosity is assumed to be 10 times the optical
luminosity, $\nu L_\nu (5100 \AA)$ (e.g., Elvis et al.\ 1994). 
[We note that up to $\sim 50\%$ of the optical luminosity 
may originate in the host galaxies (Surace \ea 2001; Bentz \ea 2006).
Thus, discussions below overestimate $\Mdot$ by up to 
a factor of $\sim 2$.]
The radiation efficiency (with respect to the rest--mass energy 
of the infalling gas) is set to be 0.05 
(e.g., Shakura \& Sunyaev 1973; Novikov \& Thorne 1973). 
Then, $\Mdot$ are estimated to be 1.3 and 5.2\,$\Msun/$yr, 
resulting in the SFR/$\Mdot$ ratios of 
2.5 and 1.8 for IRAS~04312 and IRAS~05262, respectively.

When $\Mdot$ 
exceeds the Eddington limit ($\sim 16 \, \Ledd / c^2$), however, 
the radiation efficiency drops 
due to the onset of photon trapping (Begelman 1978; Abramowicz \ea 1988).
Thus, estimates using 
accretion disk models are more adequate than those using 
a constant efficiency 
(see e.g., Kawaguchi 2003; Collin \& Kawaguchi 2004).
Here, $\Mdot$ is then estimated based 
on optical luminosity and the (standard) accretion disk model 
with a face-on disk geometry assumed (e.g., 
Bechtold \ea 1987)\footnote{The photon trapping effect takes place 
only in the inner part of the disk where UV and soft X-ray photons
emerge.
In outer region
of the disk which radiates optical continuum emission, 
self-gravity (rather than the gravity from
the central BH) governs the disk dynamics (Kawaguchi \ea 2004a).
Although the standard accretion disk model (Shakura \& Sunyaev 1973)
works only inside the self-gravity radius, 
we here adopt the standard disk model beyond the self-gravity radius 
for the sake of its simplicity.}:
\begin{eqnarray*}
 \brfrac{\Mdot}{\Msun \, {\rm yr}^{-1}} = \hspace*{6cm}
\end{eqnarray*}
\vspace{-3mm}
\begin{equation}
  2.1 \, \brfrac{\Mbh}{10^7 \Msun}^{-1}
  \, \left[
   \frac{\nu L_\nu (5100 \AA)}{10^{44}\, {\rm erg \, s^{-1}}}
  \right]^{1.5}.
 \label{eq:Mdot}
\end{equation}
With this equation, we get 19 and 76\,$\Msun\,{\rm yr}^{-1}$ for
IRAS~04312 and IRAS~05262, respectively.
Here, we obtain yet smaller SFR/$\Mdot$ ratios of 
0.17 and 0.12, 
respectively.

Then, let us discuss these ratios in the context of 
$\Mbulge - \Mbh$ relation. 
The ratio of bulge masses to BH masses in galaxies is 
about 700 (H\"{a}ring \& Rix 2004).
This number is considerably larger than 
the SFR--to--$\Mdot$ ratios for 
the two NLS1s ($\sim 1$) estimated above.
What does this huge difference mean then?

Our result (the SFR--to--$\Mdot$ ratios $\approx$ 
1 for the two NLS1s) can indicate that $\Mdot$ in NLS1s 
is temporarily large. 
To compensate the intense BH growth during the NLS1s, 
a short-term, intense starburst (with much higher SFR than the 
current SFR) happens before or in the future (i.e., 
bulge growth and BH growth phases appear without an overlapped period).
If the duration of star formation is much 
longer than the duration of AGN phase instead,
such an intense starburst is not required.

\subsection{CO Width as Surrogate for $\sigma_*$? \label{COwidth}}

Heckman \ea (1989) showed that CO and [O III] lines 
have similar widths.
While, [O III] line widths are now often used as surrogate for
bulge stellar dispersion $\sigma_*$ of AGNs (Nelson \& Whittle 1996; 
Nelson 2000; Shields \ea 2003; Boroson 2003). 
Then, Shields \ea (2006) proposed CO width as surrogate for
$\sigma_*$ (see also Ho 2007), 
although they found a systematic difference 
($\sigma_{\rm [O III]} / \sigma_{\rm CO} \approx 0.15$\,dex, on average, 
for the seven PG quasars they looked at).
By using the CO width for high redshift quasars, 
they examined the evolution of $M - \sigma_*$ relation 
at $z > 3$, arguing that high-$z$ quasars have 
smaller $M_{\rm bulge} / \Mbh$ ratios than 
quasars at $z < 3$.

Here, we compare CO width and [O III] width using our 
results, in order to assess the relation
between CO width and [O III] width for NLS1s.
Full-Width at Zero-Intensity (FWZI),
evaluated with the ranges drawn in 
figures \ref{fig:ir043spec}--\ref{fig:ir052spec} with
``flux'' labels, for the two 
NLS1s are 384\,km\,s$^{-1}$ (IRAS~04312) and 388\,km\,s$^{-1}$ (IRAS~05262).
Following Shields \ea (2006), we also 
assume that FWHM(CO) $\approx \frac{2}{3}$\,FWZI. 
Then, 
FWHM(CO) for these two IRAS galaxies are %expected to be 
about 260\,km\,s$^{-1}$, which is 
less than their ${\rm FWHM([O III])}$ widths 
(380\,km\,s$^{-1}$ and 365\,km\,s$^{-1}$, respectively). 
This is the same trend that Shields \ea (2006) presented 
for PG quasars.
Therefore, usage of CO width to estimate $\sigma_*$
can lead a systematic error upon $\Mbulge$ for NLS1s as well.
The difference between the CO and [O III] widths are
$0.15-0.17$\,dex for the two NLS1s, providing 
a factor of 4 less $\Mbulge$ for CO-based mass. 
%If [O III] widths are used as surrogate for $\sigma_*$ 
%in eq.~\ref{eq:bulmass}, 
%$\MH2 / \Mbulge$ ratios discussed in \S~6 decrease by 
%a factor of 4.

\subsection{Summary}

The tight correlation between mass of central BHs 
and mass of galactic spheroid components 
(galactic bulge or elliptical galaxy itself)
in galaxies and AGNs indicates
contemporaneous coevolution of the two massive systems.
However, no direct evidence for simultaneous growth has been found.
We aim to reveal whether 
BH growth and bulge growth take place
(quasi-) simultaneously, 
by determining the amount and distribution of molecular gas
(fuel for the coming star formation) in young AGNs.

NLS1s are characterized by relatively small BH masses
and high Eddington ratios.
Larger $\Mdot / \Mbh$ ratios of NLS1s than those of BLS1s 
indicate that BHs in NLS1s are now growing rapidly,
and that NLS1s are younger than BLS1s.
If the growth phases of BHs and bulges 
are really overlapped, 
bulges of NLS1s may also be rapidly growing with 
a significant amount of 
molecular gas accumulated in their bulges.
However, NLS1s are only 10--30\,\% among Seyfert 1 galaxies, and 
thus we do not have many NLS1s, whose CO emission is detected, 
in the nearby universe.
In order to 
investigate the nature of molecular gas in NLS1s in general, 
it is fundamental to increase 
the number of nearby NLS1s with their CO emission lines detected.

Thus, we performed 
${}^{12}$CO ($1 \rightarrow 0$) observations 
using the NMA D-configuration 
for two FIR-bright NLS1s located at $z = 0.020 - 0.032$
with 5--6'' (2--4 kpc at their distance) resolution. 
We detect their CO emission for the first time,
with $\MH2$ of $(1-3) \times 10^9 \Msun$.
These molecular--gas masses are the 2nd and 4th largest masses
 among NLS1s.

By estimating $M_{\rm dyn}$ and $\Mbulge$ with 
the CO channel maps and CO line widths, 
we get 
$\MH2 / M_{\rm dyn}$ and 
$\MH2 / \Mbulge$ ratios of 0.25--0.35 %0.41--0.45 
and 0.13--0.33, respectively.
Taking account the star formation efficiency ($\sim 0.1$), 
the increase in $M_{\rm bulge}$ in those NLS1s in the near future 
is expected not to be a huge fraction (1--5\,\% %(1--8\,\% 
of the mass of the preexisting stars).
This result can be interpreted in some ways.
Coevolution may not be simultaneous at all:
bulges (and elliptical galaxies) were formed 
in the early universe, 
while BH growth via accretion still happens today. 
Or, the bulge growth may have finished prior to the BH growth. 
Alternatively, coevolution may 
proceed with many, occasional discrete 
events, where one coevolution event produces only a small amount
of mass growth of BHs and of bulges.

We further discuss the relation between star formation rates
and accretion rates, in the context of the 
$\Mbh$--$\Mbulge$ correlation.
The two NLS1s turned out to have 
much smaller SFR/$\Mdot$ ratios ($\approx$\,1) than 
the $\Mbulge / \Mbh$ ratios of normal and active galaxies.
This huge difference can indicate either that
bulge growth and BH growth phases appear without an overlapped period,
or that the duration of star formation is much 
larger than the duration of AGN phase.

For the two NLS1s, CO widths are less than their [O III] widths 
(by 0.15--0.17 dex), 
which is the same trend that Shields \ea (2006) presented 
for several PG quasars.
Therefore, usage of CO width to estimate $\sigma_*$
can lead a systematic error 
(a factor of 4 less masses)
upon $\Mbulge$ for NLS1s as well.

\acknowledgements
We thank the NMA staff for helping observations.
This research has made use of the NASA/IPAC Extragalactic Database (NED) 
which is operated by the Jet Propulsion Laboratory,
California Institute of Technology, under contract with the National 
Aeronautics and Space Administration. 
T.K.\ thanks the financial supports 
from a Grant-in-Aid for Scientific Research of JSPS (18840038) 
and from The Research Institute of Aoyama Gakuin University.
The Nobeyama Radio Observatory is a branch of 
the National Astronomical Observatory of Japan, 
the National Institutes of Natural Sciences (NINS).

\clearpage
\clearpage
%% -------------------- Table for apj style 
\begin{deluxetable}{lcccc}
 \tablecolumns{5}
 \tabletypesize{\footnotesize}
% \scriptsize
% \tablewidth{33pc}
 \tablewidth{0pc}
% \tablenum{1}
 \tablecaption{Log of Observations}
 \label{tab:log}
 \tablehead{
\colhead{\ \ \ \ \ Target \ \ \ \ \ \ \ } &
 \colhead{redshift} & 
 \colhead{Integration Time} &
 \colhead{Beam Size} &
 \colhead{Scale} \\
\colhead{ } &
 \colhead{ } & 
 \colhead{[hours]} &
 \colhead{[$'' \times ''$]} &
 \colhead{[kpc/1$''$]} }
\startdata
 IRAS~04312\dotfill 
  & 0.020 & 6.6 & $5.8 \times 5.2$ & 0.40 \\
 IRAS~05262\dotfill 
  & 0.032 & 7.7 & $6.1 \times 5.2$ & 0.63 \\
\enddata
\end{deluxetable}
%------------------------------
%

\clearpage
\clearpage
%% -------------------- Table for apj style 
\begin{deluxetable}{lccccc}
 \tablecolumns{6}
 \tabletypesize{\footnotesize}
 \tablewidth{0pc}
% \tablenum{2}
 \tablecaption{Summary of Various Masses}
 \label{tab:mass}
 \tablehead{
\colhead{\ \ \ \ \ Object \ \ \ \ \ \ \ } &
 \colhead{$\MH2$} & 
 \colhead{$M_{\rm dyn}$} &
 \colhead{Ratio$^a$} &
 \colhead{$\Mbulge$} & 
 \colhead{Ratio$^a$} \\
\colhead{ } &
 \colhead{[$10^9\,\Msun$]} & 
 \colhead{[$10^9\,\Msun$]} & 
 \colhead{ } &
 \colhead{[$10^9\,\Msun$]} &
 \colhead{ } }
\startdata
%IRAS~04312\dotfill & 1.0 & 2.3 & 0.45 \ \ & \ \ 8.0 & 0.13 \\
%IRAS~05262\dotfill & 2.8 & 6.8 & 0.41 \ \ & \ \ 8.4 & 0.33 \\
 IRAS~04312\dotfill & 1.0 & 4.1 & 0.25 \ \ & \ \ 8.0 & 0.13 \\
 IRAS~05262\dotfill & 2.8 & 8.0 & 0.35 \ \ & \ \ 8.4 & 0.33 \\
\enddata
\tablenotetext{a}{Ratio of the molecular gas masses $\MH2$ with 
respect to the dynamical masses $M_{\rm dyn}$ or bulge masses $\Mbulge$.}
\end{deluxetable}
%------------------------------
%

\clearpage

% Figures: captions -> figures at the end of draft: for submission (aastex)
%          figure+caption : for emulateapj
% width: 8.5cm or \textwidth

%% Figure ; IR043, binned map
% For submission (aastex)
%\centerline{\includegraphics[width=\textwidth]{f1.eps}}
% For wide figures in emulateapj style
% For submission (aastex)
%\medskip

\clearpage
%% Figure ; IR052, binned map
% For submission (aastex)
%\centerline{\includegraphics[width=8.0cm]{**.epsi}}
% For wide figures in emulateapj style
% For submission (aastex)
%\medskip

\clearpage
%% Figure ; IR043, spectrum
% For submission (aastex)
%\centerline{\includegraphics[width=8.0cm]{**.epsi}}
% For wide figures in emulateapj style
% For submission (aastex)
%\medskip

\clearpage
%% Figure ; IR052, spectra
% For submission (aastex)
%\centerline{\includegraphics[width=8.0cm]{**.epsi}}
% For wide figures in emulateapj style
% For submission (aastex)
%\medskip

\clearpage
%% Figure ; IR043, channel map
% For submission (aastex)
%\centerline{\includegraphics[width=8.0cm]{**.epsi}}
% For wide figures in emulateapj style
% For submission (aastex)
%\medskip

\clearpage
%% Figure ; IR052, channel map
% For submission (aastex)
%\centerline{\includegraphics[width=8.0cm]{**.epsi}}
% For wide figures in emulateapj style
% For submission (aastex)
%\medskip

\clearpage
%% Figure ; IR043, PV diagram
% For submission (aastex)
%\centerline{\includegraphics[width=8.0cm]{**.epsi}}
% For submission (aastex)
%\medskip

\clearpage
%% Figure ; IR052, PV diagram
% For submission (aastex)
%\centerline{\includegraphics[width=8.0cm]{**.epsi}}
% For submission (aastex)
%\medskip

%
%%%UCP%%%
%\clearpage
%\plotone{f1.eps}
%\clearpage
%\plotone{f2.eps}
%\clearpage
%\plotone{f3.eps}
%\clearpage
%\plotone{f4.eps}
%\clearpage
%\plotone{f5.eps}
%\clearpage
%\plotone{f6.eps}
%\clearpage
%\plotone{f7.eps}
%\clearpage
%\plotone{f8.eps}
%

\end{document}